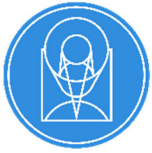

# JWST TECHNICAL REPORT

| Title:<br>Charge Migration and Residual Non-Linearity in NIRSpec BOTS Observations | Doc #: JWST-STScI-009234<br>Date: 3 November 2025<br>Rev: - |
|---|---|
| Authors:<br>Munazza K. Alam, Leonardo Ubeda, Qinyan (Apple) Lu, Néstor Espinoza, Nikolay Nikolov | Phone:<br>410-338-4445 | Release Date: 6 January 2026 |

## 1. Abstract

We investigate the effect of charge migration and residual non-linearity on the JWST/NIRSpec G395H NRS1 and NRS2 detectors using Bright Object Time Series (BOTS) observations of the ultra-hot Jupiter WASP-121b. These full-orbit phase curve observations were taken over 37.8 hours (1.57 days), and provide an excellent testbed of the non-linearity behavior of NRS1 and NRS2 over long timescales. For both detectors, our analysis demonstrates charge losses at the center of the spectral trace and charge excesses at the trace edges. We find that the NRS1 detector displays ~3× larger deviations from linearity compared to NRS2. Given the large transit (~1.5%) and eclipse (~0.5%) signals for WASP-121b, we also investigate variations in the distribution of charge throughout the time-series observation. Our results show that charge distribution varies at different planetary orbital phases for NRS1, which manifests as a change in the morphology and shape of the spectral trace over the course of the time-series. The effect of charge distribution on the trace shape is not evident for NRS2.

## 2. Introduction

Charge migration occurs when electrons move from a pixel with a lot of accumulated electrons to neighboring pixels with relatively few accumulated electrons. This migration or spilling of charge causes the "brighter fatter effect" (BFE), a nonlinear and nonlocal response that commonly occurs for photometric and spectroscopic observations of bright stars taken with the near-infrared instruments aboard the James Webb Space Telescope (e.g., Goudfrooij et al. 2024). The BFE has the most profound impact on observations in which the point spread function (PSF) is undersampled, resulting in the migration of charge from pixels in the center of the PSF to surrounding pixels. As a result, the apparent count rates deviate from incident flux in and adjacent to the PSF center.

While charge migration has been studied in detail for JWST NIRISS (e.g., Goudfrooij et al. 2024), this effect has not yet been investigated for the NIRSpec instrument. In this work, we explore the effect of charge migration and residual non-linearity (due to nonlocal effects that are beyond the scope of the linearity correction step in the `jwst` pipeline) using the NIRSpec G395H full-orbit phase curve observation of the ultra-hot Jupiter WASP-121b. This long-baseline (1.57 days) observation provides an excellent testbed for exploring charge migration and residual non-linearity in NIRSpec, since the transit and eclipse signals for this planet are so large (~1.5% transit depth; ~0.5% eclipse depth) that any changes in the charge distribution can be easily measured.

An outline of this work is as follows. In Section 3, we describe the NIRSpec G395H observations and data reduction procedures (Section 3.1), as well as our analysis methods (Section 3.2). In Section 4, we discuss our findings and summarize our conclusions in Section 5.

## 3. Data analysis

### 3.1 NIRSpec G395H Observations & Data Reduction

For this study, we used the full-orbit phase curve observations of WASP-121b taken as part of GO 1729 (PI: Mikal-Evans) and published in Mikal-Evans et al., 2023. This exposure consists of 3504 integrations taken over 1.57 days using 42 groups per integration. The observations were obtained with the NIRSpec G395H grating, the F290LP filter, and the S1600A1 slit in Bright Object Time Series (BOTS) observing mode. The SUB2048 subarray was used, which is 2048 pixels long by 32 pixels wide.

We reduced the time-series observations using the STScI `jwst` pipeline (Bushouse et al. 2025) and the open-source python-based `transitspectroscopy` package (Espinoza 2022). In Stage 1, we began with the raw uncalibrated JWST data products and applied the standard Stage 1 steps of the `jwst` pipeline (version 1.18.0, context map 1364) for time-series observations. These steps included corrections for saturation, superbias, linearity, and dark current. We applied the jump step with the default detection threshold and performed standard ramp fitting. We then used the `transitspectroscopy` package to fit the spectral trace by cross-correlating each column with a Gaussian function to find the trace position, removed outliers using a median filter with a 10-pixel-wide window, and smoothed the trace with a spline. To remove $1/f$ noise, we masked all pixels within 10 pixels from the center of the trace and subtracted the median value of all remaining pixels in each column. To extract the 1D stellar spectra, we used a 10-pixel wide aperture and summed the flux across the full wavelength range of the detector to obtain the time-series, removing 5σ outliers with a rolling median filter. We also traced the *x*- and *y*-pixel position shifts on the detector over the course of the time-series observation by cross-correlating the extracted 1D stellar spectra. We calculated the 1D wavelength map using the `assign_wcs` step. An example 2D spectrum for WASP-121b is shown in Figure 1.

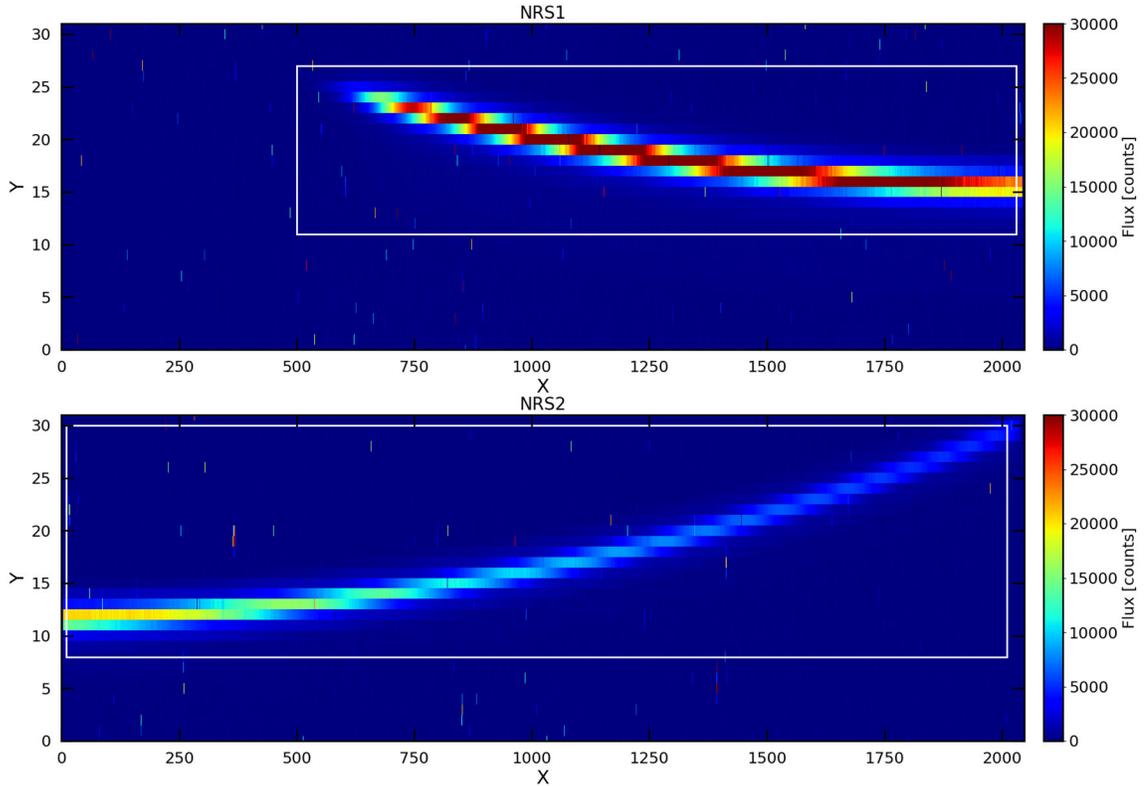

**Figure 1: Example WASP-121b spectral trace for NRS1 (top) and NRS2 (bottom) after applying the standard Stage 1 steps of the jwst pipeline. The white rectangle represents the region of the detector that was used in this study.**

### 3.2 Investigating Charge Migration & Residual Detector Non-Linearity

To measure detector linearity across all 42 groups for each of the 3504 integrations in the time-series, we restricted our analysis to the region defined by the white rectangle shown in Figure 1, which is roughly centered on the NRS1 and NRS2 spectral traces. In our analysis, we only included pixels with data quality values of zero (DQ = 0) within this selected region of interest in all 42 groups, i.e., pixels that are not marked as hot, dead, bad, low quantum efficiency, do not use, saturated, etc. by the `jwst` pipeline.

Accumulated charge in each pixel as a function of group number is shown in Figure 2. We then fit a linear function to the data using the first ten groups. We selected the first ten groups for the line fitting based on the number of groups that are necessary to reach ~10,000 counts. This 10,000-count threshold is consistent with the methods outlined in Goudfrooij et al., 2024. The linear fits for each pixel are also shown in Figure 2 as navy lines.

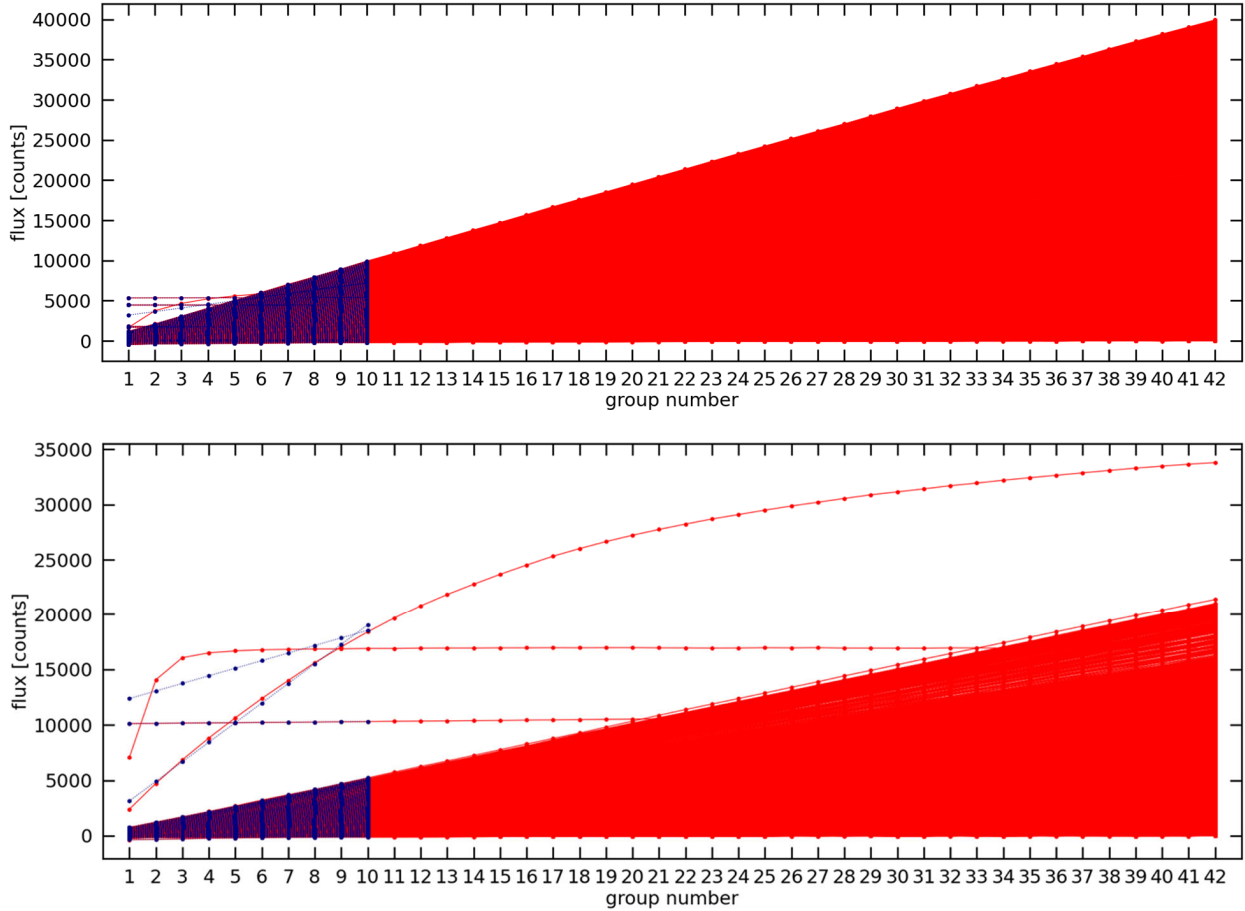

**Figure 2:** The flux of each pixel (in units of counts) as a function of group number for a given integration (red) for NRS1 (top) and NRS2 (bottom). The blue lines represent linear fits using only the first ten groups.

We then subtracted the linear fits from each individual line (i.e., for a given pixel) to investigate how the detector behaves in terms of residual non-linearity (i.e., after a linearity correction has been applied in the `jwst` pipeline). Figure 3 shows the charge difference as a function of group number. As expected, we see a reasonable fit within the first 10 groups for most pixels. Beyond the tenth group, however, some pixels show an excess of charge (positive charge difference) while others show less charge than expected (negative charge difference). We also note that there are roughly as many charge excesses as there are charge losses for pixels with residual linearity deviations, suggesting that charges that migrate are conserved. To further explore these deviations from linearity, we fit lines to the charge difference as a function of group number for each pixel with groups greater than or equal to 11. We performed tests varying the number of groups used for the linear fits and the results are consistent with charge excesses – as well as charge losses – beyond the tenth group for both NRS1 and NRS2.

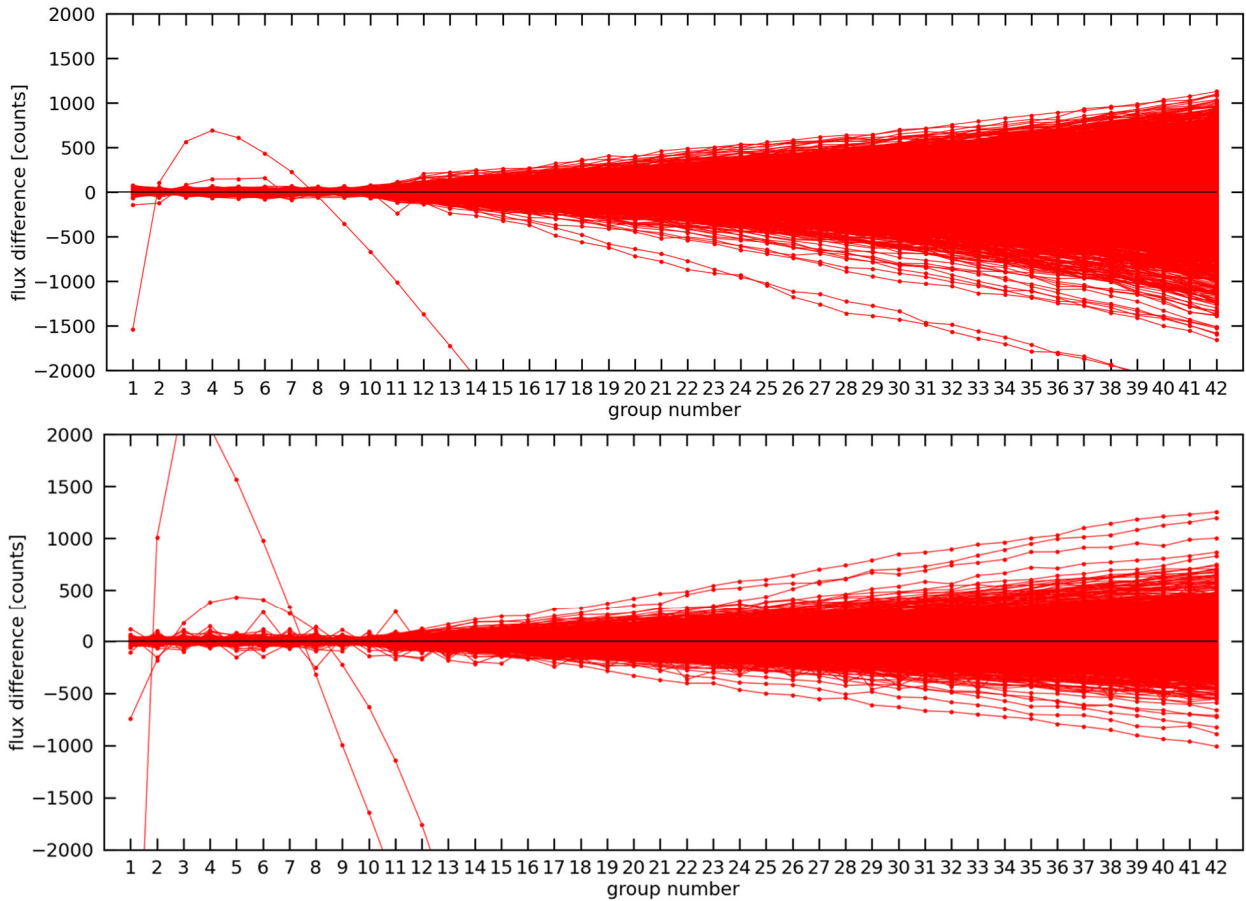

**Figure 3: Charge difference as a function of group number showing both an excess of charge (positive charge difference) and a lack of charge (negative charge difference) for NRS1 (top) and NRS2 (bottom).**

From the linear fits to each analyzed pixel, we mapped the values of the fitted charge difference slopes in arrays of size 2048 pixels long by 32 pixels wide as shown in Figure 4. Based on these pixel slope maps, we find that the largest deviations from linearity due to charge losses (i.e., negative slopes) occur for pixels located at the center of the spectral trace, whereas the largest deviations from linearity due to charge excesses (i.e., positive slopes) occur toward the edge of the spectral trace for both the NRS1 and NRS2 detectors. If a Gaussian profile was fitted to each column in the pixel slope maps, the center region of the Gaussian would correspond to a negative slope, whereas the wings of the Gaussian profile would correspond to a positive slope. The "banding" seen in the pixel slope map in Figure 4 is due to the undersampled nature of the PSF for NIRSpec BOTS observations due to the curvature of the spectral trace (see Figure 14 and Appendix B of Espinoza et al., 2023 for more details).

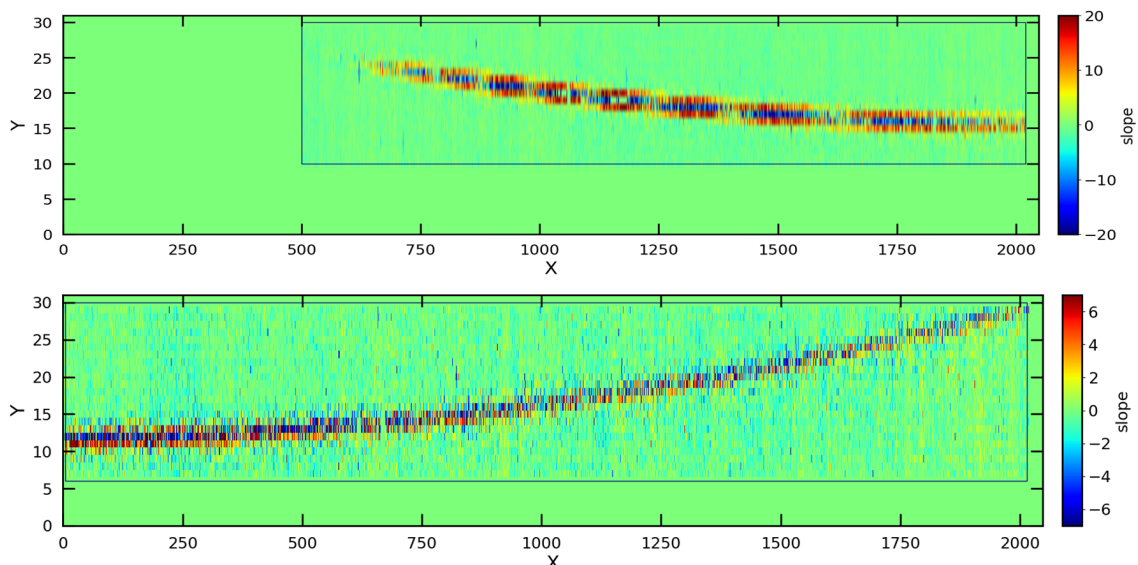

**Figure 4:** Example pixel slope map for the fitted lines to the data in the non-linearity regime (i.e., with group number greater or equal to eleven) for NRS1 (top) and NRS2 (bottom).

We also explored variations in the distribution of charge at different planetary orbital phases. To do this, we selected 100 integrations each taken in-transit (integrations 1750-1850) and in-eclipse (integrations 300-400 for the first eclipse and 2130-3230 for the second eclipse), as shown in Figure 5, and computed the mean pixel slope maps for the transit, as well as the two eclipses, as shown in Figure 6. We then created slope difference maps to compare the charge difference slopes as a function of planet phase (Figure 7).

From this exercise, we find that the distribution of charge on the NRS1 detector varies at different planetary orbital phases because charge migration depends on contrast between adjacent pixels and hence on input flux. As shown in the top panel of Figure 7, the difference of the mean pixel slope maps taken in-transit and in-eclipse reveals charge excesses (i.e., positive slopes) along the bottom of the spectral trace and charge losses (i.e., negative slopes) along the top of the spectral trace. This structure in the charge difference slopes might suggest that the morphology of the NRS1 trace is changing over the course of the time-series. This effect may be due to persistence, although further investigation regarding this behavior at the detector level is needed. The slope difference map of the first and second eclipses (bottom panel of Figure 7) does not reveal any evident structure, demonstrating that the charge distribution does not vary between the eclipses. Since the eclipse depths are comparable for the first and second eclipses (i.e., the fluxes are similar at these orbital phases), we find – as expected – that they have comparable charge distributions and thus no structure in the slope difference maps.

For NRS2, the distribution of charge does not vary as a function of planetary orbital phase, as demonstrated in Figure 8. The slope difference map comparing the primary transit and first eclipse, as well as the map comparing the slope difference for the two eclipses, both do not demonstrate any changes in the structure or morphology of the spectral trace over the course of the time-series.

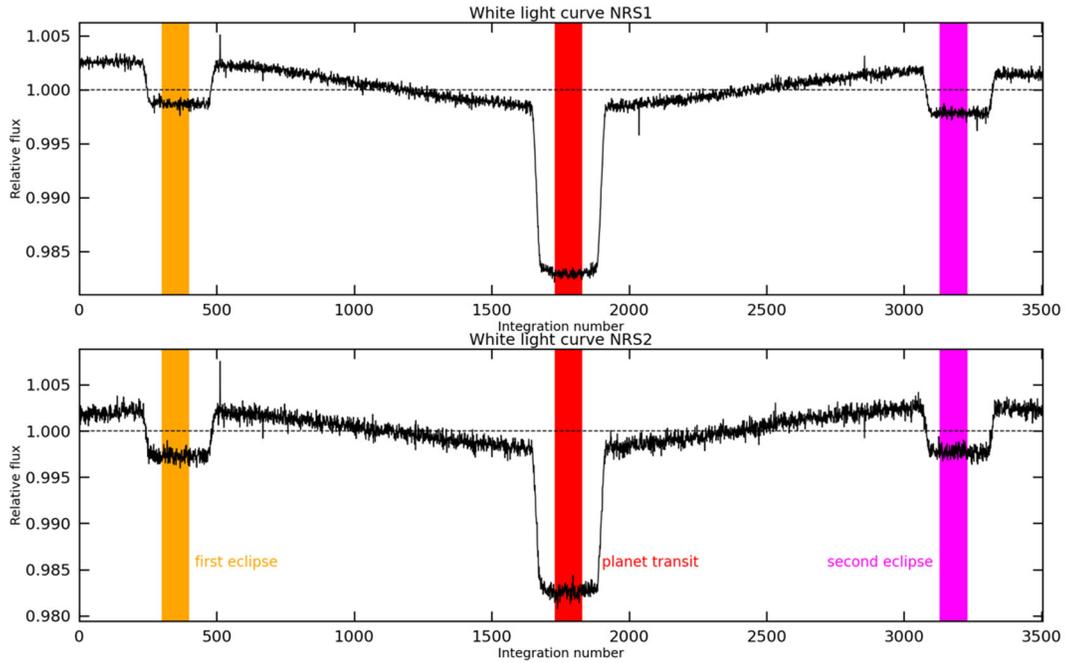

**Figure 5: Broadband full-orbit phase curve of WASP-121b NRS1 (top) and NRS2 (bottom). The primary transit (red) and two secondary eclipses (orange, magenta) are annotated. The colored shaded regions denote the integrations during the transit and eclipses that were used to compare the charge distribution at different orbital phases (see Figure 7).**

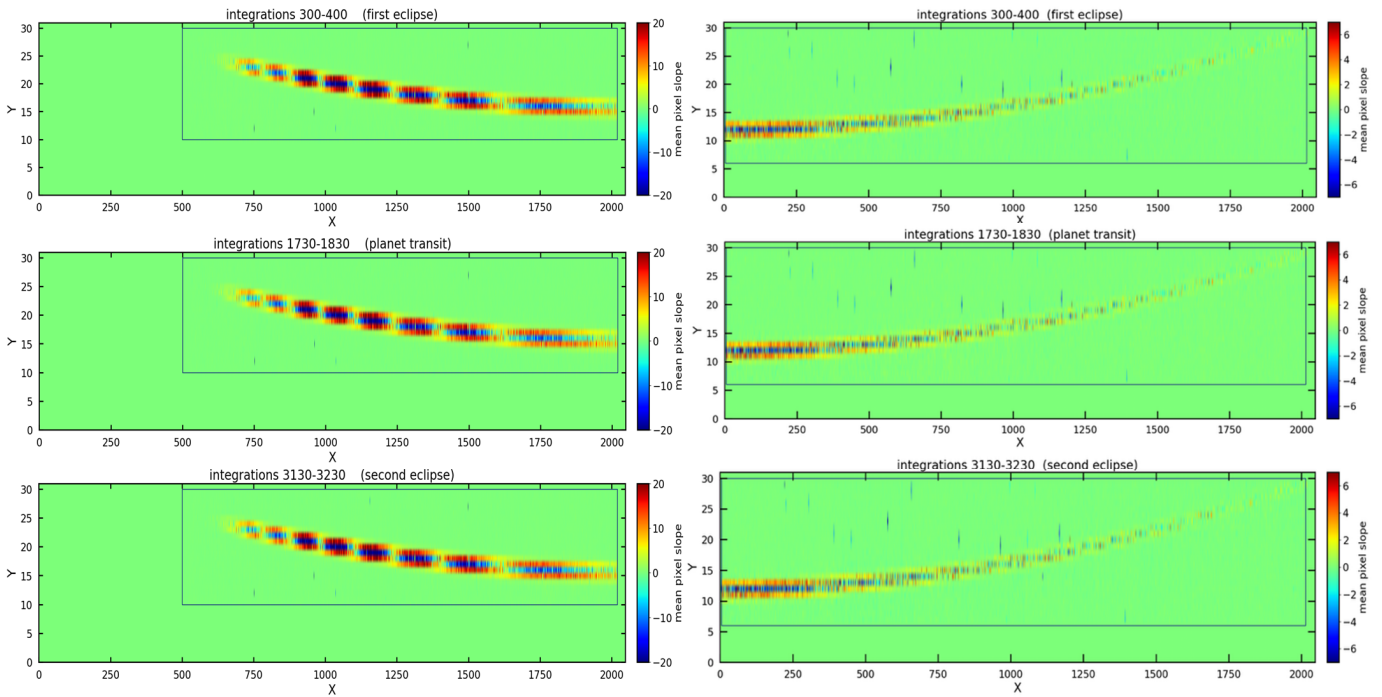

Figure 6: Mean pixel slope maps for integrations 300-400 (first eclipse; top), 1750-1850 (primary transit; middle), and 3130-3230 (second eclipse; bottom) for NRS1 (left column) and NRS2 (right column).

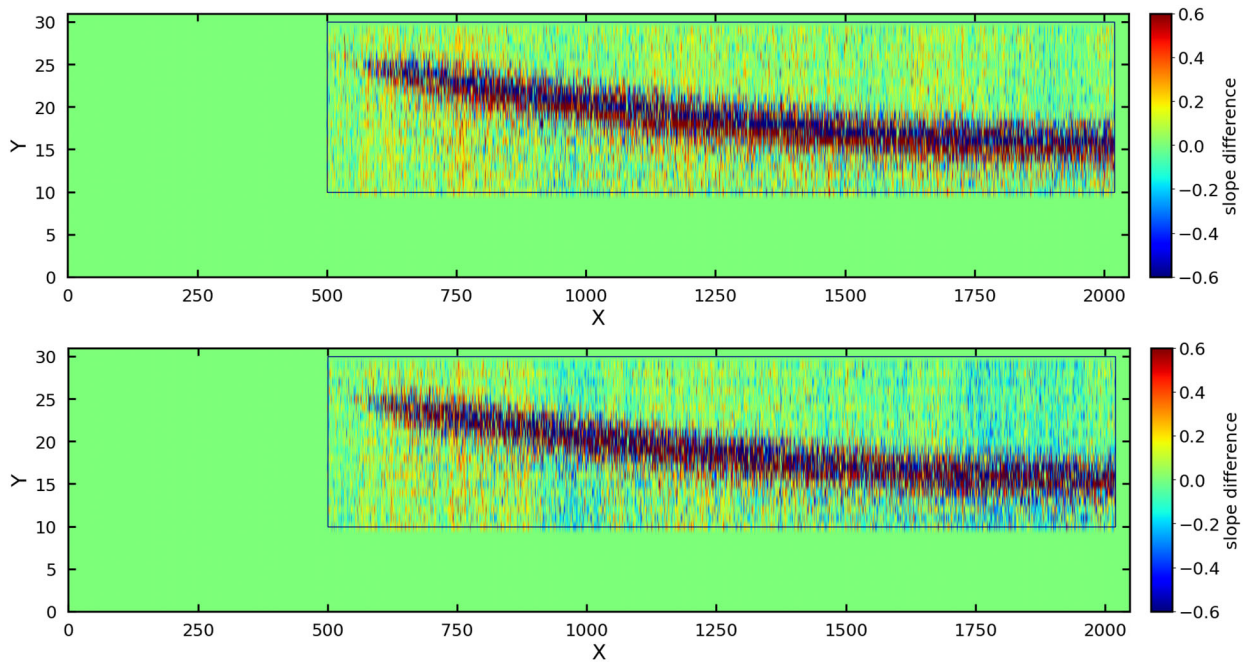

Figure 7: Top: NRS1 slope difference map created by calculating the difference between the mean pixel slope map of the primary transit and the mean pixel slope map of the first secondary eclipse. Bottom: Same as the top panel, but comparing the difference of the mean pixel slope maps of the two secondary eclipses.

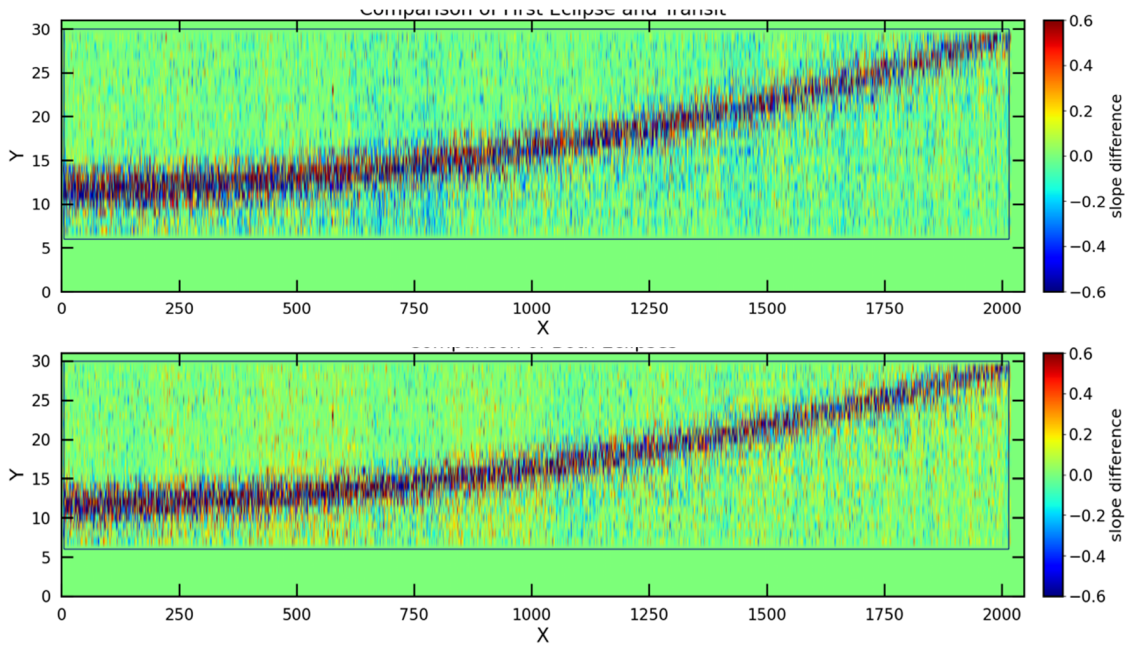

**Figure 8:** Same as Figure 7, but for NRS2.

## 4. Discussion

The NRS1 and NRS2 pixel slope maps shown in Figure 4 demonstrates the residual errors on the `jwst` pipeline linearity correction and their impact on NIRSpec BOTS observations. The NRS1 charge difference slopes range from ±20 counts/group, whereas the slopes for NRS2 show linearity deviations ranging from ±6 counts/group. It is also interesting to note that the NRS1 charge excesses (red banding) in Figure 4 are not aligned with the charge losses (blue banding). This banding is not seen in NRS2. Further, we find that the migration of charge is concentrated and seems to occur across a few (~2-3) pixels. Assuming a single aperture size for the data reduction and spectral extraction of the time-series could produce biases in the 1D stellar spectra as well as in the transit and eclipse spectra. Based on this work, we caution users against using small (<5 pixel half-width) extraction apertures for BOTS observations considering slit losses.

For time-series observations of exoplanets, residual non-linearity due to nonlocal charge migration may affect the precisions of transit depth measurements. Regan & Bergeron 2023 showed that small (~1%) errors in the non-linearity correction result in small errors on the order of ~10 ppm in the measured transit depths – but these conclusions were derived from box-shaped transit simulations (i.e., not accounting for the effect of limb darkening on transit events). However, a similar study using NIRISS SOSS data considering pixel fluences and measured error budgets on non-linearity gives transit depth offsets up to 100 ppm (Roy et al. 2025, in prep.). For NIRSpec, similarly large (~100 ppm) transit depth differences have been seen between the NRS1 and NRS2 for several high-resolution (i.e., G395H) transit observations. These offsets may be due to errors in the linearity correction, although further investigation is needed. The methodology that we follow here can be a useful prescription for exploring and quantifying the effects of charge migration in NIRSpec BOTS observations of targets across a range of stellar magnitudes.

## 5. Conclusions

We analyzed the 1.5-day NIRSpec/BOTS G395H full-orbit phase curve of WASP-121b to investigate the effects of residual non-linearity due to nonlocal charge migration on the NRS1 and NRS2 detectors. We find charge losses and excesses – indicative of charge migration – beyond the first 10 groups. In this residual non-linearity regime, charge losses are evident in the center of the NRS1 and NRS2 spectral traces, whereas charge excesses are seen at the edges of the trace. We also find that the distribution of charge varies at different planetary orbital phases for NRS1 only, which results in a change in the morphology of the trace over the course of the time-series. A follow-up study comparing the influence of charge migration and residual non-linearity behaviors on transit depth measurements between NRS1 and NRS2 for multiple targets with a range of host star magnitudes is forthcoming.